\documentclass[a4paper,11pt,oneside]{article}

\usepackage[english]{babel}
\usepackage[T1]{fontenc}
\usepackage[utf8]{inputenc}

\usepackage[pdftex,unicode]{hyperref}
\hypersetup{pdftitle=The Impact of TV Advertising on Website Traffic}
\hypersetup{pdfauthor=Lukáš Veverka and Vladimír Holý}

\usepackage{color}
\definecolor{myblue}{rgb}{0.13,0.39,0.61}
\definecolor{mygreen}{rgb}{0.24,0.68,0.64}
\definecolor{myred}{rgb}{0.93,0.33,0.23}
\hypersetup{colorlinks=true, linkcolor=myblue, anchorcolor=myblue, citecolor=myblue, filecolor=myblue, urlcolor=myblue}

\usepackage[margin=60pt]{geometry}
\usepackage{amsmath}
\usepackage{amssymb}
\usepackage{bm}
\usepackage{xfrac}

\usepackage[group-minimum-digits=3]{siunitx}
\usepackage{pifont}
\newcommand{\cmark}{\color{mygreen} \ding{51}}%
\newcommand{\xmark}{\color{myred}\ding{55}}%

\usepackage{enumitem}

\usepackage{graphicx}
\usepackage{float}
\usepackage{hhline}
\usepackage{multirow}
\usepackage{rotating}
\usepackage{booktabs}

\usepackage[authoryear]{natbib}

\begin{document}

\begin{center}
{\Large \bfseries The Impact of TV Advertising on Website Traffic}
\end{center}

\begin{center}
{\bfseries Lukáš Veverka} \\
Prague University of Economics and Business \\
Winston Churchill Square 1938/4, 130 67 Prague 3, Czech Republic \\
\href{mailto:lukas.veverka@vse.cz}{lukas.veverka@vse.cz}
\end{center}

\begin{center}
{\bfseries Vladimír Holý} \\
Prague University of Economics and Business \\
Winston Churchill Square 1938/4, 130 67 Prague 3, Czech Republic \\
\href{mailto:vladimir.holy@vse.cz}{vladimir.holy@vse.cz}
\end{center}

\begin{center}
{\itshape February 8, 2024}
\end{center}

\noindent
\textbf{Abstract:}
We propose a modeling procedure for estimating immediate responses to TV ads and evaluating the factors influencing their size. First, we capture diurnal and seasonal patterns of website visits using the kernel smoothing method. Second, we estimate a gradual increase in website visits after an ad using the maximum likelihood method. Third, we analyze the non-linear dependence of the estimated increase in website visits on characteristics of the ads using the random forest method. The proposed methodology is applied to a dataset containing minute-by-minute organic website visits and detailed characteristics of TV ads for an e-commerce company in 2019. The results show that people are indeed willing to switch between screens and multitask. Moreover, the time of the day, the TV channel, and the advertising motive play a great role in the impact of the ads.
\\

\\

\\

\section{Introduction}
\label{sec:intro}

The moment when people look up information on the internet after watching a TV ad is an excellent opportunity to shape their decision and later to convince them to finish a purchase. Since simultaneous consumption of television and internet has been confirmed and has rapidly increased, it is natural to ask how these two media types cooperate.

Recent studies that focus on immediate online response to TV ads are those of \cite{Lewis2013}, \cite{Kitts2014}, \cite{Liaukonyte2015}, \cite{Fossen2017a}, and \cite{Du2019}. However, all these studies have their distinct limitations. We identify several key components of immediate response analyses and summarize the related literature in Table \ref{tab:lit}. These include modeling and adjusting for a diurnal/seasonal pattern (in contrast to omitting the temporal aspect), modeling and estimating a gradual increase and decrease in online response by a response function (in contrast to assuming a single jump over a given time interval), allowing for each ad to have individual impact on online response (in contrast to making online response be driven purely by ad characteristics), relating online response to features of ads (in contrast to ommiting ad features), and using a nonlinear specification for the diurnal/seasonal pattern and the response-features relation (in contrast to assuming a linear specification). As you can see in Table \ref{tab:lit}, none of these studies checks all the boxes.

Our study aims to incorporate all the key components and combine the best of the previous studies. Additionally, thanks to the data available to us, we are the first to use the minute-level data about organic website traffic, which are not biased by any online paid promotion.  We separately estimate the increased number of webpage visits for each TV ad and then analyze the factors driving the success of the TV ads. We also assume the relation between the website visits and the TV ads as well as the temporal effects to be nonlinear. Specifically, we conduct our analysis in three phases: (i) the estimation of diurnal and seasonal patterns of the website visits, (ii) the estimation of a gradual increase followed by a gradual decrease in the website visits after each ad, and (iii) the estimation of dependence between the additional website visits and the ads characteristics. In line with the related literature, our overall findings support the idea of media multitasking. On the contrary with the above mentioned studies, we use the kernel smoothing method for estimation of the diurnal pattern, which allows for any kind of sudden changes in the time series and therefore prevents biased estimation of the immediate effects. Moreover, we model the response to ads by the density function of the Weibull distribution. Finally, as we do not have any assumptions about the relation between the ad effect and ad qualities, we use the random forest method.


\begin{table}
\begin{center}
\caption{The overview of the literature on immediate online responses to TV ads.}
\label{tab:lit}
\resizebox{\textwidth}{!}{
\begin{tabular}{llcccccc}
\toprule
 &  & Time & Diurnal & Response & Individual & Ad & Nonlinear \\
Research & Data & Unit & Pattern & Function & Ad Effects & Features & Relations \\
\midrule
\cite{Lewis2013}       & Online search & 1 min & \xmark & \xmark & \xmark & \xmark & \xmark \\
\cite{Kitts2014}       & Web traffic   & 5 min & \xmark & \cmark & \cmark & \xmark & \xmark \\
\cite{Liaukonyte2015}  & Web traffic   & 2 min & \cmark & \xmark & \xmark & \xmark & \xmark \\ 
\cite{Fossen2017a}     & Online WOM    & 2 min & \xmark & \xmark & \cmark & \cmark & \xmark \\
\cite{Du2019}          & Online search & 1 min & \cmark & \xmark & \xmark & \cmark & \xmark \\ 
\textbf{This Research} & Web traffic   & 1 min & \cmark & \cmark & \cmark & \cmark & \cmark \\
\bottomrule
\end{tabular}
}
\end{center}
\end{table}

\section{Literature Review}
\label{sec:lit}

Media multitasking during TV ads is becoming more and more common. The research of this behavior can provide a valuable insight into advertising (see, e.g., \citealp{Duff2019}). The fact that people do not like ads that much was confirmed e.g.\ by \cite{Wilbur2016} who estimated that 27 \% of commercial breaks are interrupted by switching TV channels. As for the interaction with TV ads, several studies suggest that the cross-media effects should be taken into account as there is a significant increase in online search and shopping following TV advertising (see, e.g., \citealp{Joo2014}, \citealp{Lewis2013}, \citealp{Du2019}). Especially the informational content of an ad increases online search (see \citealp{Chandrasekaran2017a}). Naturally, ads perform better when the product relates to the context (see \citealp{Guitart2017}). The highest effectiveness in terms of conversions is reached when the ads are placed on the station where there are no competitors (see \citealp{Guitart2020}). TV is able to drive not only searches but also social activity. The short-term positive effect on online chatter is stronger in the visibility and virality rather than the popularity of a brand (see Tirunillai and Tellis\cite{Tirunillai2017}). There is also an impact on the number of tags (both brand and TV channel) caused by TV ads in social interactivity with TV (see \citealp{Fossen2017a}). Moreover, the higher the cumulative usage of social networks, the higher the shopping activity (see \citealp{Zhang2017}). \cite{Fossen2019} then discovered that social chatter increases shopping behaviour as a result of increased ad responsiveness.

The research most relevant to our topic shows that TV ads lead to online responses such as brand website traffic (see \citealp{Kitts2014}, \citealp{Liaukonyte2015}), online word of mouth (WOM) (see \citealp{Fossen2017a}), and online search (see \citealp{Lewis2013}, \citealp{Du2019}). The studies stem from various types of analyses using monthly, weekly, daily, hourly, or minute-level data. To our knowledge, \cite{Lewis2013} published the first case study examining the immediate response to TV on a dataset at the minute level. In that study, there is also the first indication of a possible response function showing how much time people need to search for more information on their phones. The response function was estimated by \cite{Kitts2014} as the log-normal distribution. Even though the suggested methodology was capable of working with minute data, the study was performed on the aggregated data in 5 minute time chunks. \cite{Liaukonyte2015} as well as \cite{Kitts2014} have proven that TV ads increase not only online search but also the website traffic by including the brand website traffic in the model. \cite{Fossen2017a} then included for the first time the ad features (e.g. the position of a spot in a break) in the analysis. \cite{Du2019} extended the study by \cite{Liaukonyte2015} by focusing on the effects of ad features (e.g.\ creative, slot of the break, time). It is included in the model as a response rate where all ad characteristics have linear dependencies. Both \cite{Liaukonyte2015} and \cite{Du2019} decompose diurnal and seasonal trends of webpage visits using fixed effects representing weeks in a year, days of the week, hours of the day, and minutes of the hour. \cite{Liaukonyte2015} do not include interaction between these variables, which results in the loss of ability to distinguish the effect of diurnality and seasonality except for the top-level effects for weeks. Additionally, it does not allow to estimate the constantly increasing webpage visits in a certain hour and constantly decreasing webpage visits in another hour. \cite{Du2019} then include a within-hour trend that may vary by hour of the week. However, this utilizes a significant number of degrees of freedom. In addition, it does not cover an unexpected behaviour such as an increase during special occasions (e.g.\ offers, sales). The key characteristics of the reviewed studies are compared in Table \ref{tab:lit}.

\section{Analyzed Data Sample}
\label{sec:data}

Our dataset covers website traffic and TV ads details of a fast-moving consumer goods (FMCG) company, which sells their products online only, in the year 2019. The company operates in Europe with focus on the region of Central and Eastern Europe. It is a well-known company which belongs among the market leaders\footnote{In accordance with confidentiality agreements, the company name has been kept anonymous.}. The data contains a time series of organic website visits obtained from Google with the minute frequency resulting in the length of $n=$ 525,600 minutes, i.e.\ 365 days. Since the webpage visits are only organic, we do not need to consider effects of on-line advertisement. In Google Analytics, ``organic search'' encompasses unpaid traffic resulting from natural search engine rankings. Users find a website through non-paid listings on search engine results pages, and this traffic is automatically tracked by Google Analytics without the need for additional parameters. On the other hand, ``paid search'' denotes traffic generated through paid advertising, such as Google Ads, where users click on sponsored listings. Even though it might be interesting to examine interaction of TV and on-line communication, it is beyond the scope of this paper.

While analyzing traffic sources in Google Analytics, it's important to acknowledge a potential limitation. Users might be more inclined to click on organic search results based on the content they observe at the top of the listings, particularly where sponsored links are displayed. However, the existence and significance of this phenomenon are debatable and extend beyond the scope of our current research. Although we don't delve into investigating this behavior, recognizing its potential impact serves as a caveat and acknowledges a limitation in the scope of our model.

Furthermore, we have information for $m=3,222$ TV ads which contains the date and the time of each broadcasting with an accuracy of one second and three categorical variables describing qualities of each spot:
\begin{itemize}
\item An advertising motive consisting of 1 sponsoring (carrying a branding message) and 11 regular promotional spots (9 communicates discounts, the other 2 special offers).
\item A premium position -- first, second, other, penultimate, and last.
\item A TV channel where the TV Channel 2 is the most popular local TV channel. TV Channels 3, 4, and 5 are in the same bundle as the TV Channel 2, however, target slightly different audience and have lower viewership. The audience of TV Channel 6 are older people and finally TV Channel 7 broadcasts music videos.
\end{itemize}

\section{Diurnal and Seasonal Adjustment of Website Visits}
\label{sec:diurnal}

\subsection{Motivation}

\label{sec:diurnalMotiv}

\begin{figure}
\begin{center}
\includegraphics[width=0.9\textwidth]{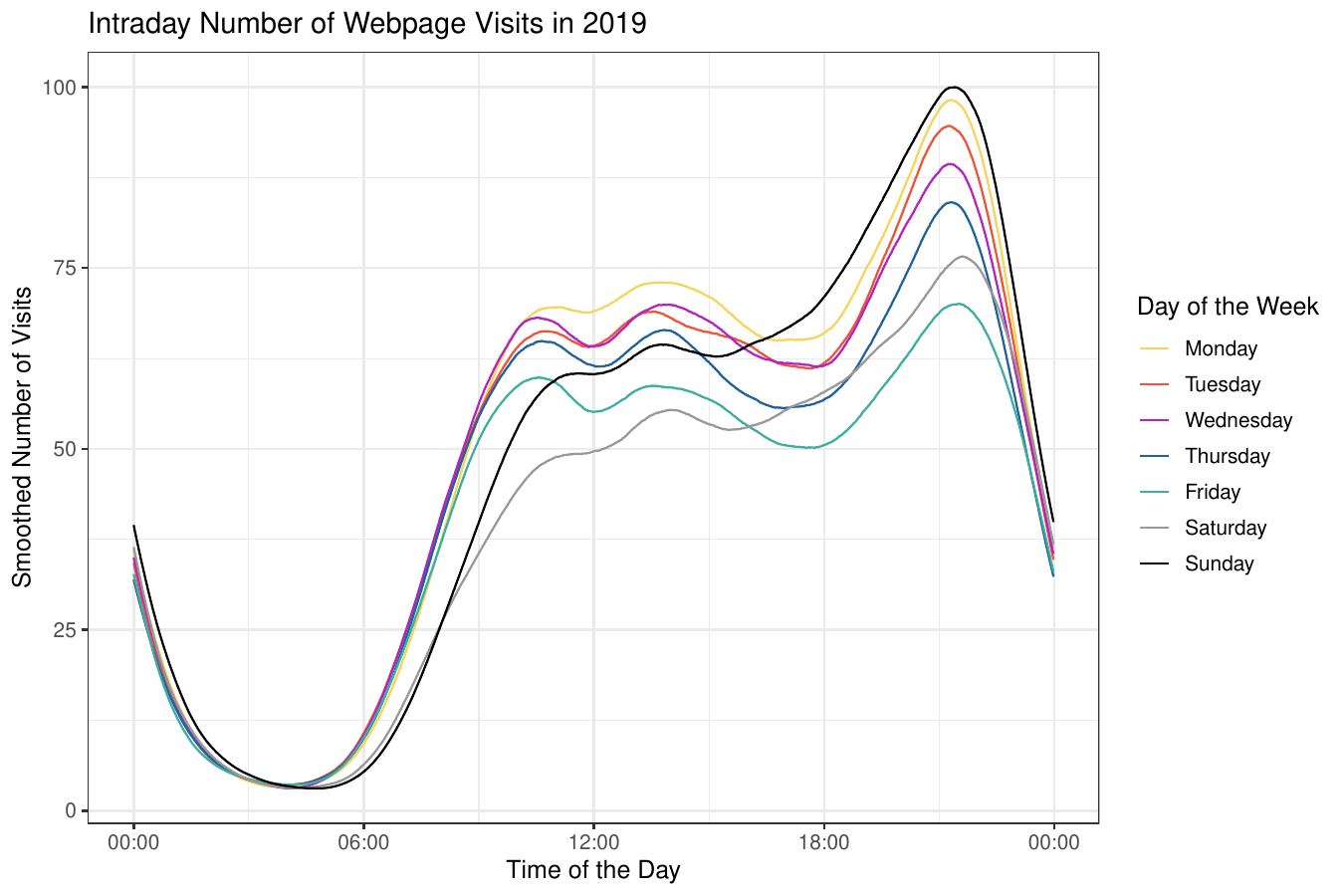}
\caption{The intraday number of website visits smoothed by the Epanechnikov kernel with bandwidth 60. Data are taken from 2019.}
\label{fig:diurnal}
\end{center}
\end{figure}

\begin{figure}
\begin{center}
\includegraphics[width=0.9\textwidth]{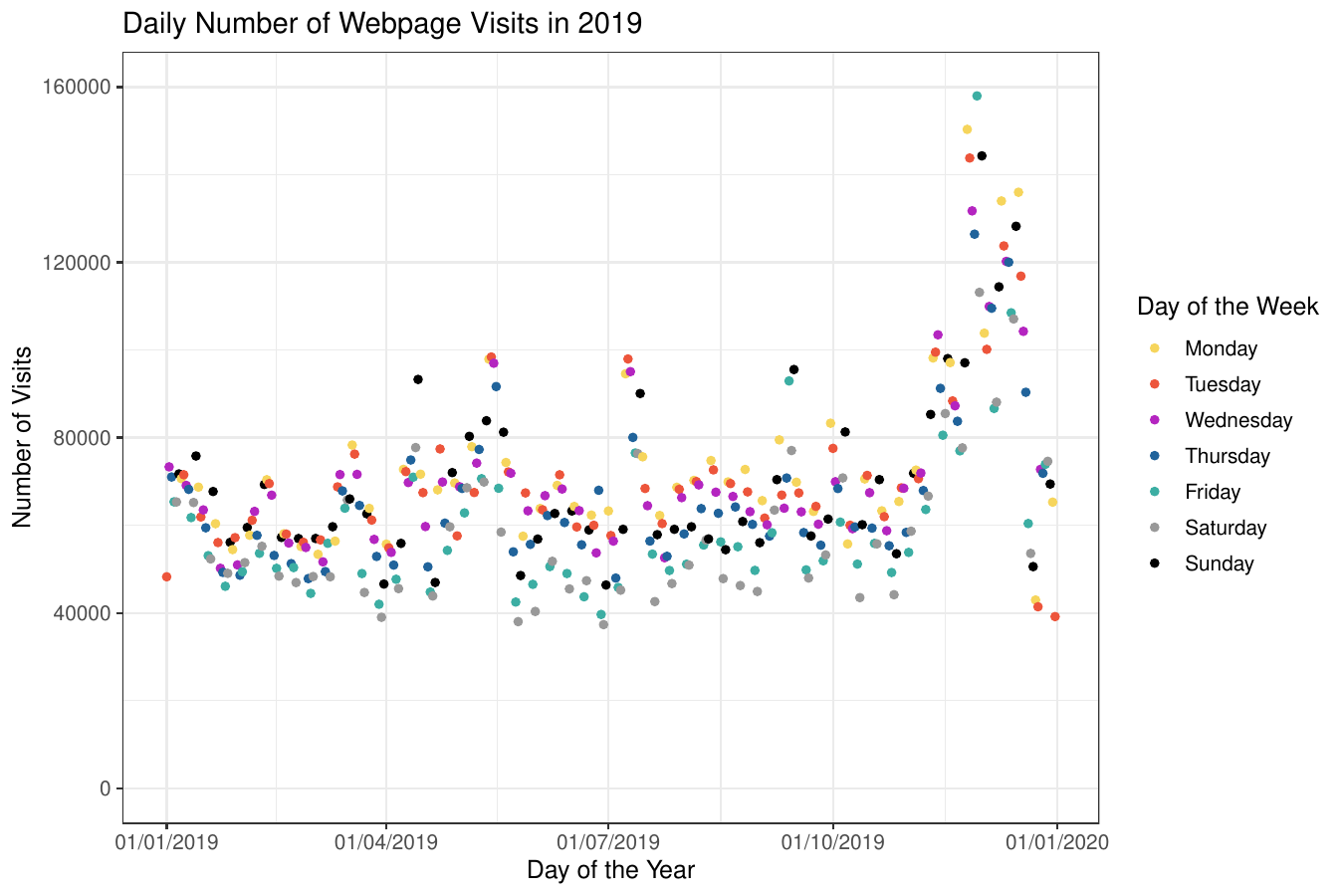}
\caption{The daily number of website visits. Data are taken from 2019.}
\label{fig:year}
\end{center}
\end{figure}

Let us study intraday and daily behavior of website visits. Figure \ref{fig:diurnal} shows the smoothed number of website visits during a day. There is a strong pattern common for all days of the week with the maximum around 21:30 and the minimum between 3:00 and 5:00. The course of the individual days of the week, however, slightly differs. For example, the highest traffic in the afternoon is on Mondays while the highest traffic in the evening is on Sundays. The lowest traffic in the afternoon is on Saturdays while the lowest traffic in the evening is on Fridays. Of course, not all Fridays are the same. Figure \ref{fig:year} shows the total number of daily website visits for the individual days of the year 2019. On average, Fridays together with Saturdays have the lowest traffic. Nevertheless, the highest daily traffic is on Friday. Not surprisingly, it is the Black Friday which is the busiest shopping day of the year in many countries in recent years. The takeaway from these two figures is that website visits have strong but rather complex diurnal and seasonal patterns.

The goal of this section is to adjust website visits for these patterns. Note that we do not study the causes of diurnality and seasonality, nor do we forecast website visits. Both these tasks are undoubtedly interesting but beyond the scope of this paper. We simply aim to determine the expected number of website visits at a given time in history using adjacent observations. Subsequently, this allows us to study deviations of the observed website visits from their expected values.

\subsection{Methodology}
\label{sec:diurnalMeth}

To estimate the expected number of website visits, we use a kernel smoother. It is a non-parametric method which locally averages observations. At this point, let us assume that there are no ads on TV promoting the website. Let $Y_t \in \mathbb{N}_0$ denote the random variable representing the number of visits at time $t=1,\ldots,n$ and $y_t \in \mathbb{N}_0$ its observed value. The expected value $\lambda_t$ is then estimated as
\begin{equation}
\label{eq:kernelSmooth}
\hat{\lambda}_t = \frac{\sum_{k = -h}^h K \left(\frac{k}{h + 1} \right) y_{t+k} }{\sum_{k = -h}^h K \left(\frac{k}{h + 1} \right)},
\end{equation}
where $K(u)$ is a kernel function with support on $[-1, 1]$ and $h$ is a bandwidth for the local averaging. Note that we restrict ourselves to kernels with bounded support as it greatly facilitates the computation in our case.

We consider several commonly used kernel functions:
\begin{itemize}
\item The triangular kernel, $K(u) = 1 - |u|$, $u \in [-1, 1]$.
\item The Epanechnikov kernel, $K(u) = \frac{3}{4} \left(1 - u^2 \right)$, $u \in [-1, 1]$.
\item The quartic kernel, $K(u) = \frac{15}{16} \left(1 - u^2 \right)^2$, $u \in [-1, 1]$.
\item The triweight kernel, $K(u) = \frac{35}{32} \left(1 - u^2 \right)^3$, $u \in [-1, 1]$.
\item The tricube kernel, $K(u) = \frac{70}{81} \left(1 - |u|^3 \right)^3$, $u \in [-1, 1]$.
\end{itemize}

We select the optimal kernel function $K(u)$ and the optimal bandwidth parameter $h$ using repeated 2-fold cross-validation. We randomly split the sample to a training subsample and a validation subsample of equal sizes. We use the observations from the training subsample to estimate the expected values \eqref{eq:kernelSmooth} for the validation subsample. We assess the fit based on the average log-likelihood of the Poisson distribution
\begin{equation}
\frac{1}{\lvert{\mathcal{V}}\rvert} \sum_{t \in \mathcal{V}} y_t \ln(\lambda_t) - \lambda_t - \ln(y_t!),
\end{equation}
where $\mathcal{V}$ denotes the set of indices in the validation subsample. The average log-likelihood is non-positive and larger values indicate better fit. Although we do not require any distributional assumptions on $Y_t$ at this point, we choose this measure as we assume Poisson distribution for $Y_t$ in Section \ref{sec:resp}. We repeat this procedure 1,000 times and report average values.

\subsection{Results}
\label{sec:diurnalRes}

\begin{figure}
\begin{center}
\includegraphics[width=0.9\textwidth]{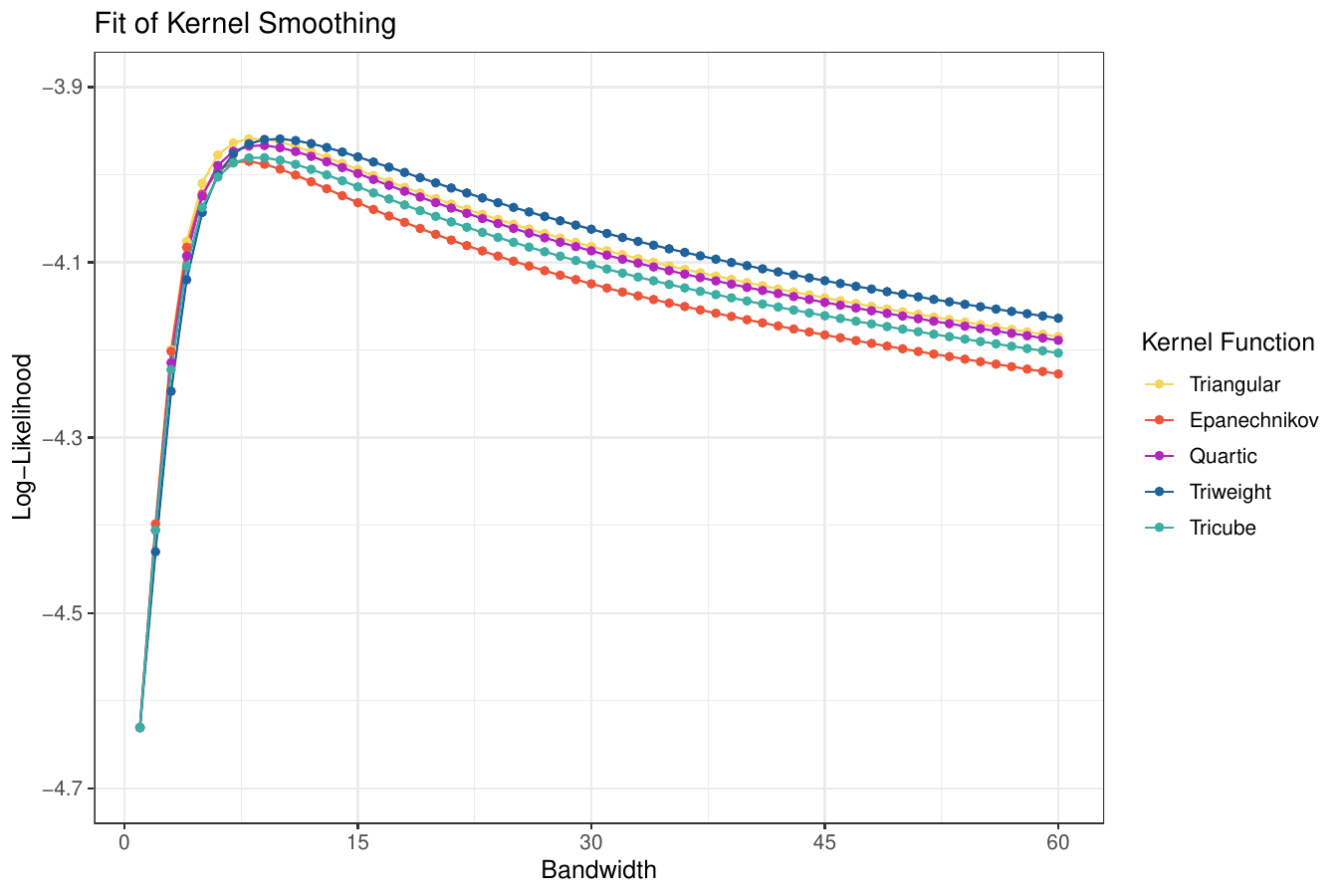}
\caption{The average log-likelihood of kernel smoothing based on various kernel functions and bandwidth parameters.}
\label{fig:band}
\end{center}
\end{figure}

Figure \ref{fig:band} shows the performance of the considered kernel functions with bandwidth parameter $h$ ranging from 1 to 60 minutes. We exclude observations within 30 minutes after each aired ad to remove influence of ads on website visits. Generally, smaller values of $h$ result in larger variance (i.e.\ undersmoothing) while larger values result in larger bias (i.e.\ oversmoothing). In our case, the best fit has average log-likelihood -3.959036 and is obtained for the triangular kernel with bandwidth 8. The second best fit has average log-likelihood -3.959176 and is obtained for the triweight kernel with bandwidth 10. Using the mean squared error instead of the average Poisson log-likelihood would also identify the triangular kernel with bandwidth 8 as the best choice. In the sequel, we use the triangular kernel with bandwidth 8. As seen in Figure \ref{fig:band}, the model is robust to the choice of kernel function as it differs minimally. The only problem can be caused by setting too narrow bandwidth. However, once a sufficient bandwidth is used, whether a slightly narrower or (almost) arbitrarily wider than the optimal, it leads to very similar results.

\section{Immediate Response to TV Ads}
\label{sec:resp}

\subsection{Motivation}
\label{sec:respMotiv}

\begin{figure}
\begin{center}
\includegraphics[width=0.9\textwidth]{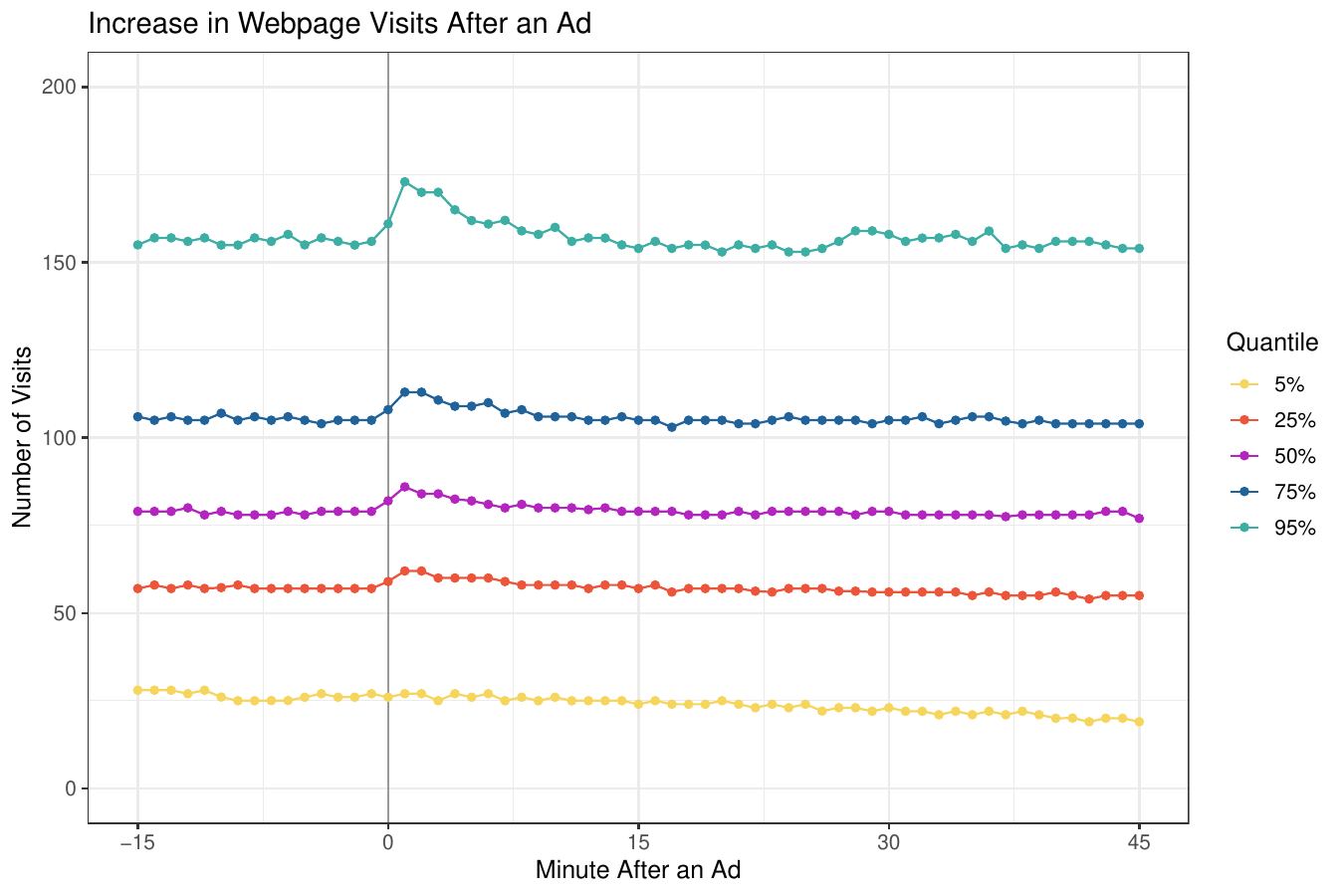}
\caption{The quantiles of website views with respect to the time before or after an aired ad.}
\label{fig:quantile}
\end{center}
\end{figure}

Let us study the influence of the individual ads on website visits. One can expect that for a few minutes after an ad is aired, there will be an increase in website visits. The question is exactly how many visits are caused by the ad and how they are distributed in time. Before we proceed to a rigorous analysis, we first take a look at Figure \ref{fig:quantile}. For each ad, we create the time series of website visits starting 15 minutes before the ad and ending 45 minutes after the ad. Figure \ref{fig:quantile} then shows quantiles of these time series for each minute. Note that we do not adjust for diurnal and seasonal patterns here and the ads are not evenly distributed in time. This leads to a modest decline in the 5\%-quantile. We can see that there is indeed an increase in website visits the first minute after the ad. In the following minutes, website visits decrease to their standard level. This is most apparent for the 95\%-percent quantile but also holds for the other quantiles except the 5\%-quantile. Therefore, most ads prompt an immediate response in terms of website visits, but some go unnoticed.

The goal of this section is to discern website visits by people who saw an ad and those who did not. We model visits unrelated to ads by the method proposed in Section \ref{sec:diurnal}. Simultaneously, we determine the number of additional website visits for each individual ad and estimate the shape of their distribution in time, which is assumed to be common for all ads. 

\subsection{Methodology}
\label{sec:respMeth}

Let us assume that the random number of website visits $Z_t \in \mathbb{N}_0$ at time $t=1,\ldots,n$ is comprised of two components -- the number of visits by people who saw an ad $X_t \in \mathbb{N}_0$ and the number of visits by those who did not $Y_t \in \mathbb{N}_0$. Furthermore, we assume that both components follow the Poisson distribution with time-varying rate, $X_t \sim \mathrm{Pois}(\mu_t)$ and $Y_t \sim \mathrm{Pois}(\lambda_t)$, respectively. This assumption is motivated by the inhomogeneous Poisson point process with the complete independence and memoryless properties, often used as the arrival process in the queueing theory. The total number of website visits then also follows the Poisson distribution, $Z_t = X_t + Y_t \sim \mathrm{Pois}(\mu_t + \lambda_t)$. The probability mass function is given by
\begin{equation}
\label{eq:poisProb}
\mathrm{P} \left[ Z_t = z \mid \mu_t, \lambda_t \right ] = \frac{(\mu_t + \lambda_t)^z e^{-(\mu_t + \lambda_t)}}{z!}.
\end{equation}
The expected value and the variance are simply $\mathrm{E}[Z_t] = \mathrm{var}[Z_t] = \mu_t + \lambda_t$.

We do not observe the two individual components but only their sum $z_t \in  \mathbb{N}_0$, $t=1,\ldots,n$. Additionally, we observe the times of ads $s_j \in \mathbb{R}^+$, $j=1,\ldots,m$. Note that while the website visits are recorded at discrete times $t \in \mathbb{N}^+$ (in our case minutes), we allow the times of ads to be recorded with higher precision (in our case seconds) and we treat them as real numbers. Our goal is to estimate the rates $\mu_t$ and $\lambda_t$, $t=1,\ldots,n$. For this purpose, let us impose some assumptions on their structure. 

\begin{figure}
\begin{center}
\includegraphics[width=0.9\textwidth]{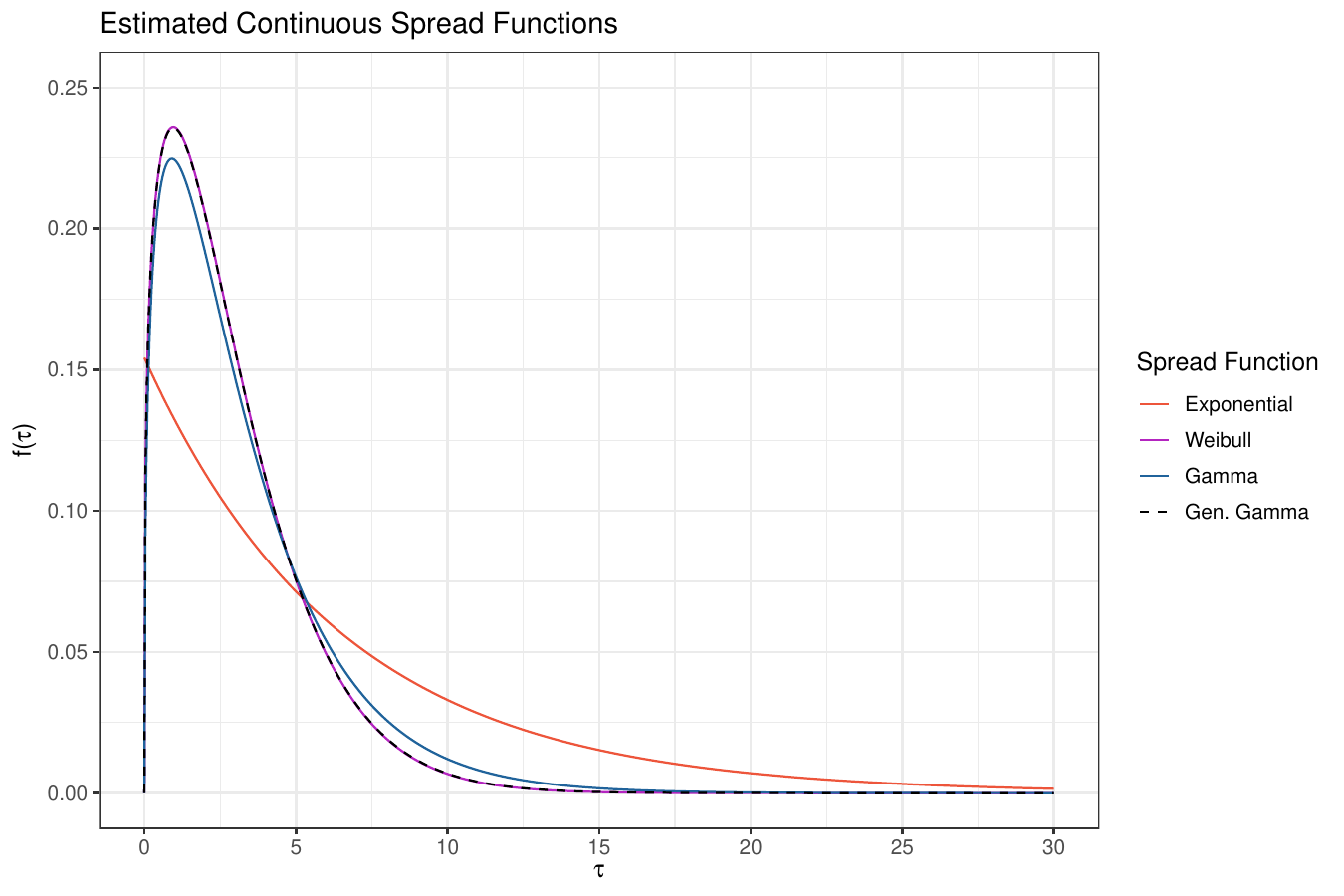}
\caption{The exponential, Weibull, gamma, and generalized gamma spread functions with the parameters estimated by the maximum likelihood method. The Weibull spread function overlaps with the generalized gamma spread function.}
\label{fig:spread}
\end{center}
\end{figure}

First, we focus on people who saw an ad and visited the website afterwards. Naturally, there is a delay between the end of an ad and a subsequent website visit. For visits caused by an ad that ends at time $s \in \mathbb{R}^+$, we denote the probability of a visit occurring at time $t \in \mathbb{N}^+$ as the spread function. We assume this spread function is given by
\begin{equation}
\label{eq:spread}
V_s(t) = \begin{cases}
0 & \text{for } t \leq s, \\
\int_{0}^{t - s} f(\tau) \, d \tau & \text{for } s < t \leq s + 1, \\
\int_{t - s - 1}^{t - s} f(\tau) \, d \tau & \text{for } t > s + 1, \\
\end{cases}
\end{equation}
where $f(\tau)$ is a non-negative continuous function satisfying $\int_{0}^{\infty} f(\tau) \, d \tau = 1$. For any $s \in \mathbb{R}^+$, it holds that $V_s(t) \in [0, 1]$, $t = 1, \ldots \infty$ and $\sum_{t=1}^{\infty} V_s(t) = 1$. We refer to $V_s(t)$ as the discretized spread function and $f(\tau)$ as the continuous spread function. As a candidate for $f(\tau)$, we consider the density function of the generalized gamma distribution of \cite{Stacy1962}, defined as
\begin{equation}
\label{eq:gengammaDensity}
f(\tau) = \frac{\alpha \varphi}{\Gamma \left( \psi \right) }\left( \tau \alpha \right)^{\psi \varphi - 1} e^{- \left( \tau \alpha \right)^{\varphi}} \qquad \text{for } \tau \in (0, \infty),
\end{equation}
where $\alpha$ is the scale parameter, $\varphi$ with $\psi$ are the shape parameters, and $\Gamma(\cdot)$ is the gamma function. We also consider its special cases -- the gamma distribution for $\varphi = 1$, the Weibull distribution for $\psi = 1$, and the exponential distribution for $\psi = 1$ and $\varphi = 1$. The generalized gamma distribution is quite flexible and is often used to choose a suitable distribution among parametric distributions with positive values. In the case of the generalized gamma density, the integrals in \eqref{eq:spread} have the form
\begin{equation}
\label{eq:gengammaIntegral}
\int_{\tau_1}^{\tau_2} f(\tau) \, d\tau = \frac{\Gamma \left( \psi, (\tau_1 \alpha)^{\varphi} \right) - \Gamma \left( \psi, (\tau_2 \alpha)^{\varphi} \right)}{\Gamma(\psi)} = Q \left( \psi, (\tau_1 \alpha)^{\varphi}, (\tau_2 \alpha)^{\varphi} \right),
\end{equation}
where $\Gamma(\cdot, \cdot)$ is the incomplete gamma function and $Q(\cdot,\cdot,\cdot)$ is known as the generalized regularized incomplete gamma function. The expected value is given by
\begin{equation}
\label{eq:gengammaMean}
\mathrm{E}[\mathcal{T}] = \frac{1}{\alpha} \frac{\Gamma \left( \psi + \frac{1}{\varphi} \right) }{\Gamma(\psi)}.
\end{equation}
We can interpret it as the average delay between the end of an ad and a subsequent website visit. The mode is given by
\begin{equation}
\label{eq:gengammaMode}
\mathrm{Mode}[\mathcal{T}] = \begin{cases}
\frac{1}{\alpha} \left( \psi - \frac{1}{\varphi} \right)^{\frac{1}{\varphi}} & \text{for } \varphi \psi > 1, \\
0 & \text{for } \varphi \psi \leq 1. \\
\end{cases}
\end{equation}
The generalized gamma density function is increasing for values lower than mode and decreasing for larger values. Examples of the course of the continuous spread function are shown in Figure \ref{fig:spread}.

We let each ad have a different impact on website visits and associate it with parameter $\theta_j$, $j=1,\ldots,m$. The expected number of website visits by people who saw an ad is then 
\begin{equation}
\label{eq:rateMu}
\mu_t = \sum_{j=1}^m \theta_j V_{s_j}(t).
\end{equation}
For website visits by people who did not see an ad, we use the expected value as in Section \ref{sec:diurnal} and set
\begin{equation}
\label{eq:rateLambda}
\lambda_t = \frac{\sum_{k = -h}^h K \left(\frac{k}{h + 1} \right) \left( z_{t+k} - \mu_{t+k} \right)}{\sum_{k = -h}^h K \left(\frac{k}{h + 1} \right)}.
\end{equation}
Note that, unlike in Section \ref{sec:diurnal}, we employ the full dataset and do not exclude observations immediately after ads. Vector $\mu = (\mu_1,\ldots,\mu_n)$ is a function of $\alpha, \varphi, \psi, \theta_1, \ldots, \theta_m$ while vector $\lambda = (\lambda_1,\ldots,\lambda_n)$ is a function of $z_1,\ldots,z_n,\mu_1,\ldots,\mu_n$.

We find the estimates of parameters $\alpha, \varphi, \psi, \theta_1, \ldots, \theta_m$ by maximizing the log-likelihood
\begin{equation}
\label{eq:lik}
\left( \hat{\alpha}, \hat{\varphi}, \hat{\psi}, \hat{\theta}_1, \ldots, \hat{\theta}_m \right) \in \arg \max \sum_{t = 1}^n \ln \mathrm{P} \left[ Z_t = z_t \mid \lambda_t, \mu_t \right].
\end{equation}
We solve non-linear optimization problem \eqref{eq:lik} by a numerical method. Unfortunately, this is very demanding as each of the potentially large number of ads has its own parameter $\theta_j$. In our case, there are 3,222 ads and therefore 3,225 parameters to be estimated. To tackle the optimization problem, we have to make some concessions.

We limit the impact of an ad in a distant future. We can see from \eqref{eq:spread}, \eqref{eq:gengammaDensity}, and \eqref{eq:rateMu} that each ad affects $\mu_t$ in an infinitely long future $t > s$. Although the discretized spread function $V_s(t)$ quickly approaches zero for larger values of $t$, it nevertheless remains positive. It is therefore reasonable to introduce a cut-off point $d$ from which the function is reduced to exactly zero. We redefine the discretized spread function as
\begin{equation}
\label{eq:spreadCut}
V_s(t) = \begin{cases}
0 & \text{for } t \leq s, \\
\frac{\int_{0}^{t - s} f(\tau) \, d \tau}{\int_{0}^{d - s + \lfloor s \rfloor} f(\tau) \, d \tau} & \text{for } s < t \leq s + 1, \\
\frac{\int_{t - s - 1}^{t - s} f(\tau) \, d \tau}{\int_{0}^{d - s + \lfloor s \rfloor} f(\tau) \, d \tau} & \text{for } s + 1 < t \leq s + d, \\
0 & \text{for } t > s + d. \\
\end{cases}
\end{equation}
In our application, we set $d = 30$, i.e.\ we assume that viewers visit the website within half an hour. Figure \ref{fig:spread} illustrates that this cut-off is sufficient. If there are more than $d$ minutes between two ads, they do not directly affect website visits at the same time through $\mu$. They can, however, still affect the website visits at the same time indirectly through $\lambda$, but the kernel function in \eqref{eq:rateLambda} has also a cut-off point. Therefore, if there are more than $2h+d$ minutes between two ads, they do not affect website visits at the same time at all. This provides a sort of independence between distant ads.

Let us assume for a moment that we know the optimal values of parameters $\alpha$, $\varphi$, and $\psi$. We therefore need ``only'' to estimate parameters $\theta_1, \ldots, \theta_m$. Such task can then be divided into several smaller optimization problems given our independence property. We separate the time series of ads into groups when a break of more than $2h+d$ minutes without an ad occurs. We then find the optimal values of $\theta_j$ independently for each group. In our application with bandwidth 8, we separate 3,222 ads into 665 groups with the largest group containing 84 ads, which is quite manageable.

The problem is that we cannot separate parameters $\theta_j$ when parameters $\alpha$, $\varphi$, and $\psi$ are not known. It is because the shape of the discretized spread function is estimated from the course of all ads. We resort to the block coordinate descent algorithm, in which we alternate estimation of parameters $\alpha$, $\varphi$, and $\psi$ with fixed $\theta_1, \ldots, \theta_m$ and estimation of $\theta_1, \ldots, \theta_m$ with fixed $\alpha$, $\varphi$, and $\psi$. For more details on this approach, see e.g.\ \cite{Wright2015}. For the numerical optimization in each iteration, we use the Nelder--Mead algorithm of \cite{Nelder1965}.

\begin{figure}
\begin{center}
\includegraphics[width=0.9\textwidth]{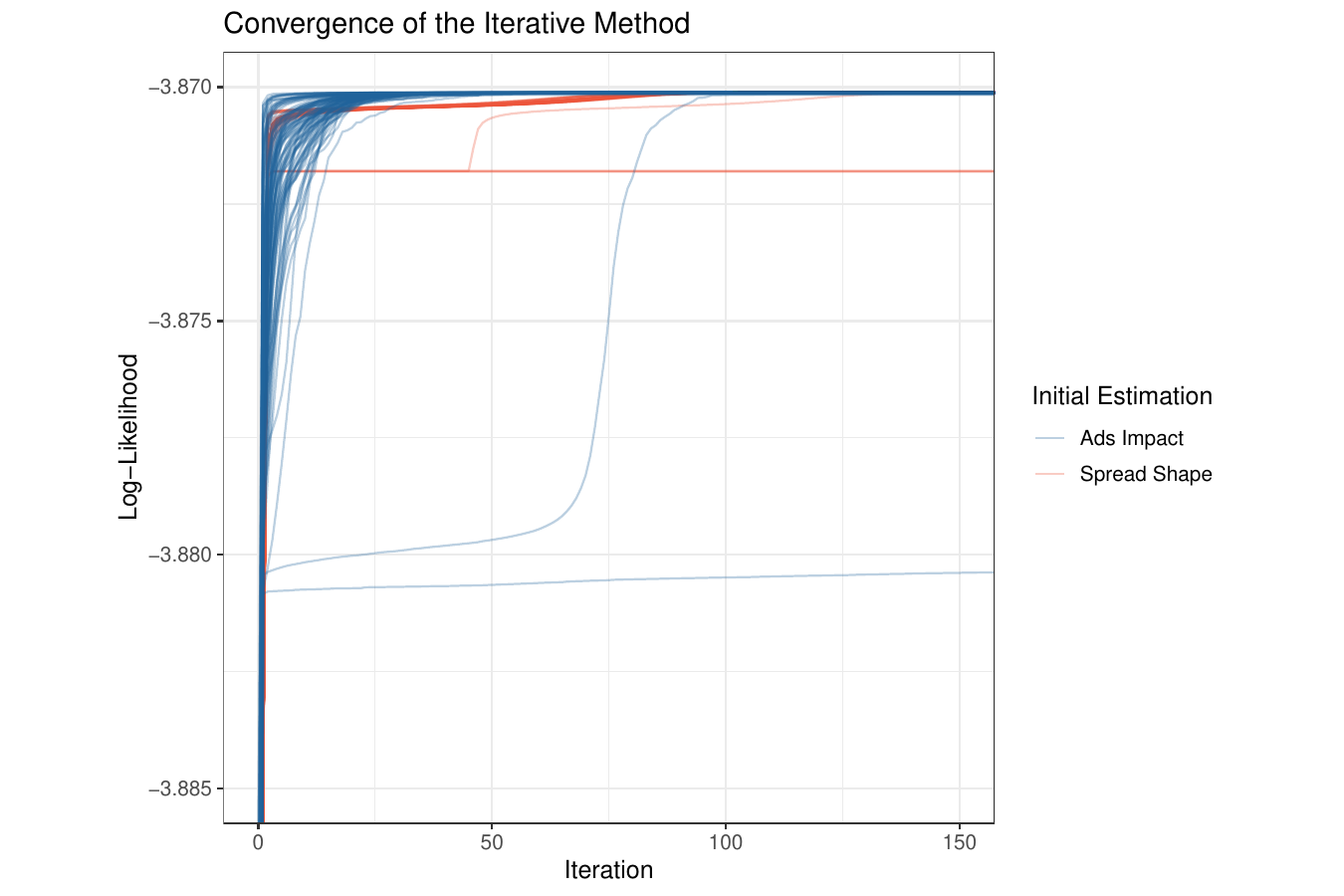}
\caption{The progress of the log-likelihood computed by the block coordinate descent algorithm with 200 random starting values.}
\label{fig:converge}
\end{center}
\end{figure}

To illustrate the performance of our approach, we conduct a simple simulation experiment, focusing on the Weibull spread function (i.e.\ setting $\psi=1$). We randomly generate starting values for the parameters $\alpha$ and $\varphi$ from the standard log-normal distribution, and for the parameters $\theta_1, \ldots, \theta_m$ from the discrete uniform distribution on the interval $[1, 100]$. Two versions of the algorithm were employed: one initiating with the estimation of ads impact, $\theta_1, \ldots, \theta_m$, and the other initiating with the estimation of the spread shape, $\alpha$, $\varphi$. For both versions, we simulate 100 sets of starting values. Figure \ref{fig:converge} shows the convergence of the estimation. In the vast majority of scenarios, the block coordinate descent algorithm together with the Nelder--Mead algorithm converge to the global maximum. However, there are a few exceptions. In three scenarios, initiating with spread shape results in being stuck in a local optimum. Interestingly, there is another scenario that follows a very similar path but is able to escape this local optimum and converge to the global optimum. In two scenarios, initiating with ads impact leads to very slow computation. One scenario was terminated before reaching convergence due to our constraints on computing time. The other scenario converged to the global optimum. In both cases, the starting values consist of small $\alpha$ (0.15, 0.13) and large $\varphi$ (7.03, 9.40), leading to a distinct shape of the spread function concentrated around a high number (6.36, 7.53) and close to zero for lower values. For reliable and fast convergence, we recommend initiating with the estimation of ads impact using reasonable starting values for the spread shape such as $\alpha=\varphi=\psi=1$ (i.e.\ the exponential distribution with unit mean).

\subsection{Results}
\label{sec:respRes}

First, we focus on the estimated continuous spread function. The estimated values of parameters $\hat{\alpha}$, $\hat{\varphi}$, and $\hat{\psi}$ together with descriptive and goodness-of-fit statistics are reported in Table \ref{tab:fit}. The shape of the estimated continuous spread function is shown in Figure \ref{fig:spread}. Overall, the Weibull, gamma, and generalized gamma spread functions perform very similarly. In terms of the average log-likelihood, the Weibull and generalized gamma spread functions have the best fit. However, based on the Wilks likelihood-ratio test, the generalized gamma spread function reduces to the Weibull spread function. The Akaike information criterion, which penalizes the number of parameters, also suggests the Weibull spread function. The estimated Weibull spread function is increasing on interval $(0, 0.95)$ and decreasing on interval $(0.95, \infty)$. This means that it takes about one minute until the website visits caused by an ad reach their peak. The estimated gamma and generalized gamma spread functions show similar behavior. This is in contrast to the exponential spread function, which can be only decreasing and is therefore not suitable. In the sequel, we use the Weibull spread function.

\begin{figure}
\begin{center}
\includegraphics[width=0.9\textwidth]{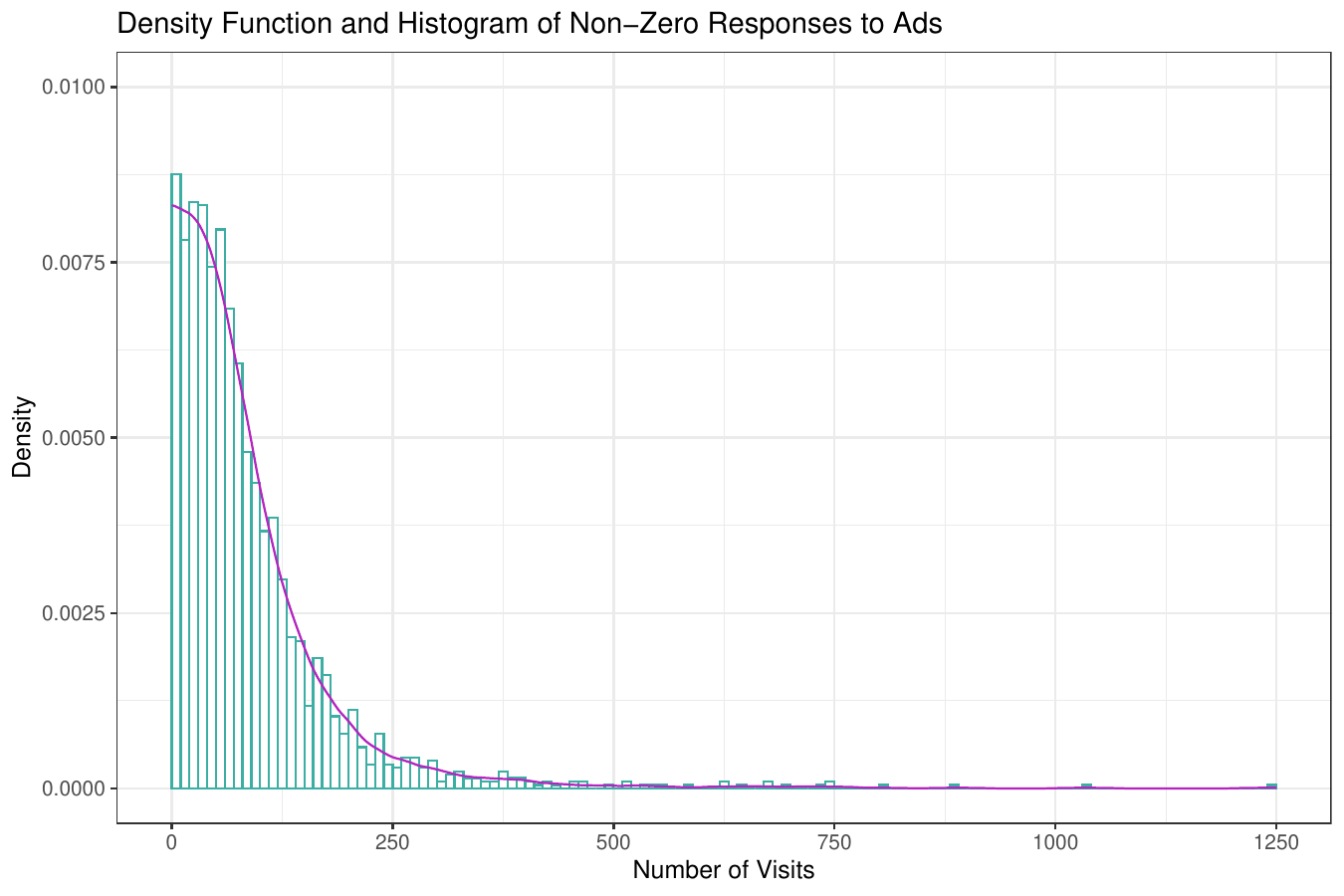}
\caption{The density function estimated by the Epanechnikov kernel with bandwidth 20 and the histogram with bin width 10 of the estimated non-zero number of additional website visits caused by the individual ads.}
\label{fig:theta}
\end{center}
\end{figure}

Second, we focus on the estimated number of additional website visits caused by the individual ads, i.e.\ on parameters $\hat{\theta}_1, \ldots, \hat{\theta}_m$. About a third of ads, specifically 1,177 out of 3,222 ads, have exactly zero impact on website visits. The distribution of the remaining two thirds is shown in Figure \ref{fig:theta}. The average number of website visits caused by an ad (when ommiting those with zero impact) is 87.09 and the maximum is 1,240.20. We further analyze the estimated values $\hat{\theta}_1, \ldots, \hat{\theta}_m$ in the following section.

\begin{table}
\begin{center}
\caption{The estimated spread function parameters, the mean, the mode, the average log-likelihood, the Akaike information criterion, and the p-value of the Wilks test for the models with the exponential, Weibull, gamma, and generalized gamma spread functions.}
\label{tab:fit}
\begin{tabular}{lcccccccc}
\toprule
Spread Function & $\hat{\alpha}$ & $\hat{\varphi}$ & $\hat{\psi}$ & Mean & Mode & Log-Lik. & AIC & Wilks \\ 
\midrule
Exponential & 0.15 & 1.00 & 1.00 & 6.49 & 0.00 & -3.870395 & 4,075,005 & 0.00 \\ 
Weibull     & 0.32 & 1.28 & 1.00 & 2.90 & 0.95 & -3.870118 & 4,074,716 & 0.97 \\ 
Gamma       & 0.42 & 1.00 & 1.38 & 3.27 & 0.91 & -3.870121 & 4,074,719 & 0.08 \\ 
Gen. Gamma  & 0.32 & 1.27 & 1.01 & 2.91 & 0.95 & -3.870118 & 4,074,718 &   -- \\ 
\bottomrule
\end{tabular}
\end{center}
\end{table}

\section{Factors Driving the Success of TV Ads}
\label{sec:factors}

\subsection{Motivation}
\label{sec:factorsMotiv}

Once the additional website visits caused by the ads $\hat{\theta}_1, \ldots, \hat{\theta}_m$ are estimated, we focus on the explanation of the effect of ads using their characteristics. For each ad, we know the TV channel of the ad, the position of the ad within the block of ads in which it was aired, and the advertising motive of the ad. As the website visits caused by the ads are likely to exhibit similar diurnal and seasonal dynamics as the unrelated visits in Figures \ref{fig:diurnal} and \ref{fig:year}, we also include a time of the day, a day in the week, and a month in the year as explanatory variables. Overall, we expect highly non-linear and complex relationships among the six variables. We therefore build a random forest to reveal the dependency of success of ads on their characteristics. Unlike in Section \ref{sec:diurnal}, our goal is to forecast the additional website visits. Based on these findings, it will be possible to determine the value of each ad slot and, consequently, optimize the selection of ad slots.

\subsection{Methodology}
\label{sec:factorsMeth}

Random forests are a machine learning method based on bootstrap aggregation of regression trees. To avoid creating an ensemble of very similar trees, each tree is allowed only a certain number of variables to be selected in a split node. This decorrelation technique leads to a robust model. The amount of variables to be selected in each node is a tuning parameter of random forests. Among the other tuning parameters belong the number of trees, the minimum size of terminal nodes, and the sample size to draw. For a textbook treatment of random forests, see e.g.\ \cite{Hastie2008}.

Random forests might be sensitive to a setting of the tuning parameters in the sense of the robustness as well as the accuracy. We set the number of trees to 10,000 and select the other tuning parameters from a grid using repeated cross-validation. The number of variables selected in a split node is considered to be 1, 2, 3, 4, 5, and 6. The minimum size of terminal nodes is considered to be 5, 10, 15, 20, 25, and 30. The sample size to draw is considered to be 50\%, 60\%, 70\%, 80\%, 90\%, and 100\%. Each combination of the tuning parameters forms one setting. Therefore, we have 216 settings of the tuning parameters in total. The cross-validation of tuning parameters of the random forest proceeds as follows. First, the data is split randomly into 3 groups -- the training sample consisting of 50\% of all observations, the validation sample of 25\%, and the testing sample of 25\%. Second, a random forest is built on the training sample for each parameter setting. Third, dependent variable $\theta_j$ is predicted and the mean square error (MSE) is calculated for all three samples. To obtain precise results, we repeat these steps 10,000 times. The parameter setting with the lowest MSE in the validation sample is then selected.

The final model is constructed on the full data sample using the best performing tuning parameters. Due to the random nature of the method, we estimate the model 10,000 times and report the average results. This rather large number of models together with equally large number of trees is chosen in order to obtain precise results, particularly for the variance importance (see, e.g., \citealp{Behnamian2017}). If we were only interested in predicted values, we could take a more computationally efficient way and use a fraction of the repetitions and trees.

As far as the interpretation of any machine learning model is concerned, it is always difficult. In the case of random forests, the variable importance plot, which reflects the amount of accuracy and/or node impurity reduction for each variable, proves to be a useful tool. Note that the highest reduction suggests the most important variable. To understand the dependencies between the ad effect and the explanatory variables, we use partial dependence plots. They can be used with any kind of machine learning model and their key principle is to visualize the average marginal effect of each variable while other variables remain constant. In other words, the plotted function at a particular value of the explanatory variable represents the average prediction if we plug this value to all observations (see, e.g., \citealp{Molnar2019}).

\subsection{Results}
\label{sec:factorsRes}

Based on the cross-validation, we set the number of selected variables to 5, the minimum size of terminal nodes to 15, and the drawn sample size to 50 \%. Table \ref{tab:treeError} shows the mean absolute error (MAE), the root mean square error (RMSE), and the coefficient of determination (R$^2$) for the random forests built in the cross-validation. As expected, all three criteria have the best values on the training sample and the worst on the test sample, although the performance on the test sample is very similar to the validation sample.

\begin{figure}
\begin{center}
\includegraphics[width=0.9\textwidth]{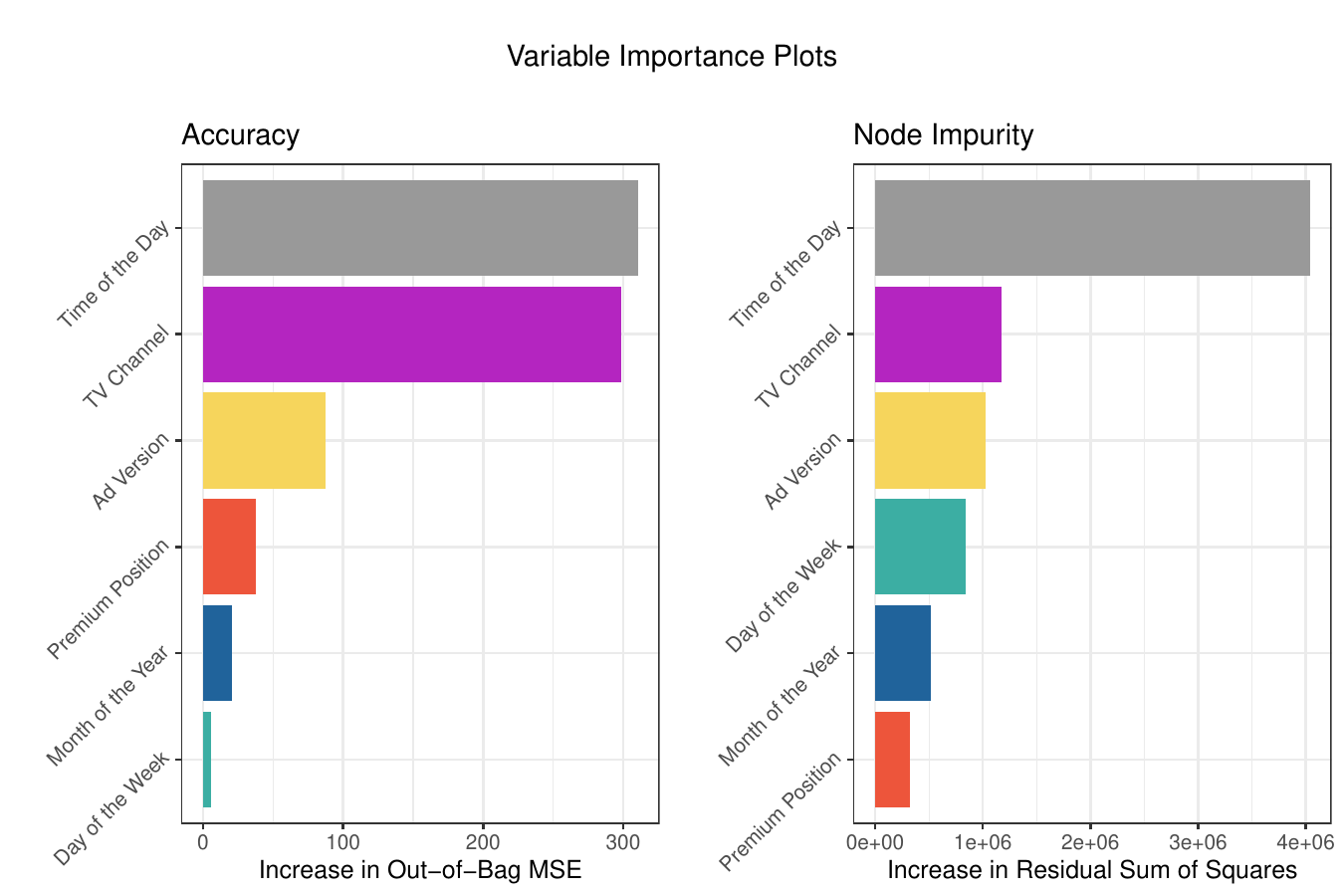}
\caption{The most important variables in terms of the accuracy and node impurity in descending order.}
\label{fig:importance}
\end{center}
\end{figure}

\begin{figure}
\begin{center}
\includegraphics[width=0.9\textwidth]{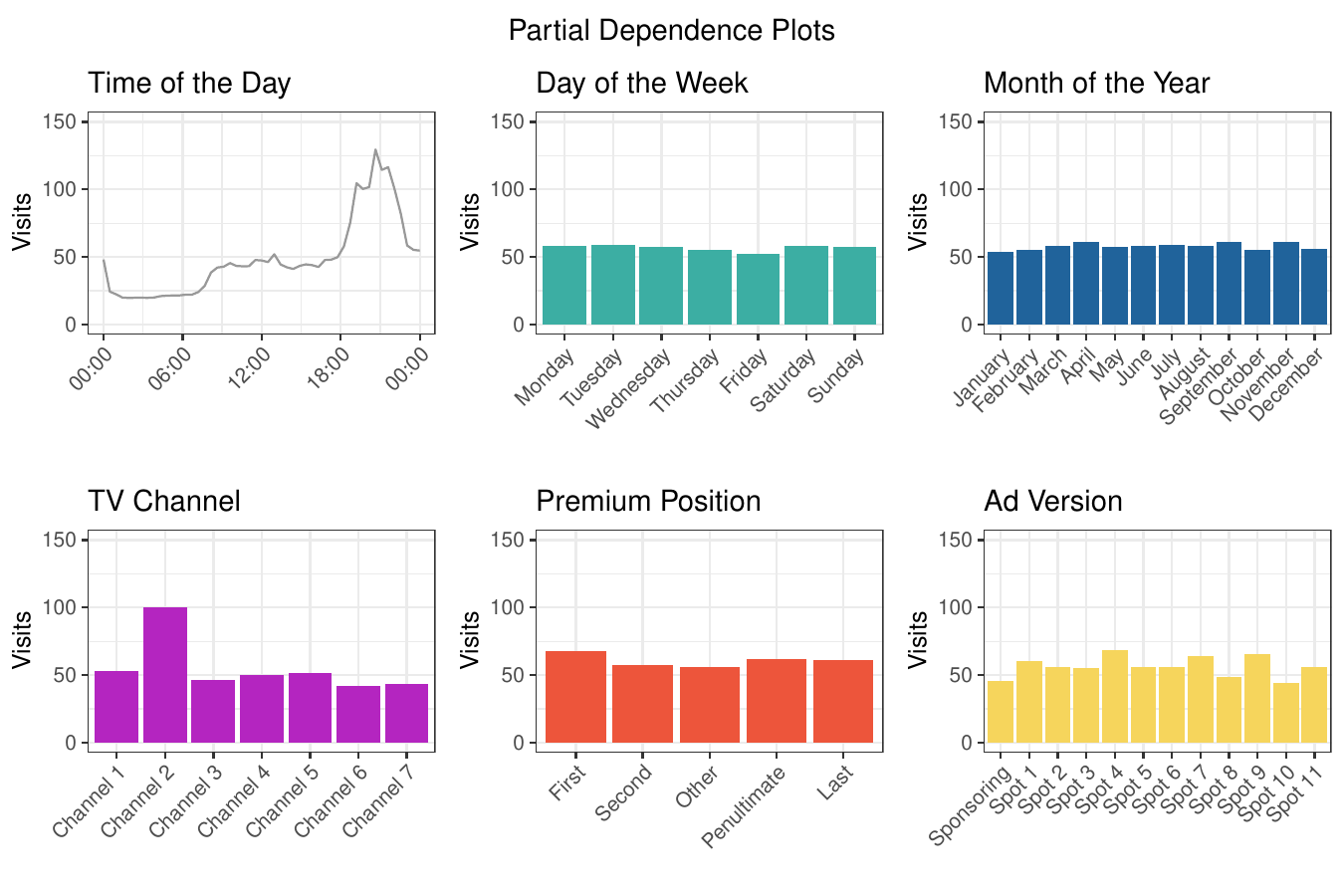}
\caption{Partial dependence plots for the six explanatory variables.}
\label{fig:partial}
\end{center}
\end{figure}

The fit of the final model is also assessed in Table \ref{tab:treeError}. The R$^2$ is 0.49 on the full sample suggesting a decent fit of the model. The prediction error of the model can be measured by comparing the observations with the values predicted by trees that were not grown using the individual observations. The out-of-bag R$^2$ then drops down to 0.21. The variable importance plots are presented in Figure \ref{fig:importance} and the partial dependence plots in Figure \ref{fig:partial}. The most important variable is the time of the day followed by the TV channel and the motive.

The partial dependence plot for the time of the day variable shows a very similar pattern to Figure \ref{fig:diurnal}. This is not at all surprising as ads aired during the prime time are more watched and, consequently, more expensive. The other two temporal variables -- day of the week and month of the year -- are less important but nevertheless in line with Figure \ref{fig:year}. For example, Fridays have the lowest additional visits.

The TV channel can heavily influence the website traffic as shown in its partial dependence plot. In our case, Channel 2 attracts by far the most visitors on average. It should be noted, however, that the buying process of ads might be limited to the choice of TV bundles instead of specific TV channels. It is then possible within our model to calculate the average effect of the channels belonging to the bundle and select the most profitable one.

The effect of the premium position variable is also quite distinguishable. The most effective position in the break is naturally the first one. After the first ad, many TV viewers change the channel. Therefore, the second and following positions are the worst. Interestingly, the penultimate position performs better than the last position. This is likely caused by people not willing to search for more information online once the TV programme is already running. For the immediate response, it is therefore not the best choice.

Finally, the advertising motive also significantly influence the website traffic. Interestingly, the sponsoring version of the ad, which is limited to the first and last positions, is the second worst in terms of attracting website visitors. This is likely connected to the fact that the motive of a sponsoring is quite restricted and cannot contain any kind of promotion such as price, sales, or product benefits. Spot 4, on the other hand, performs the best. This is not surprising as the main message of this motive is 70 percent discount -- the highest discount among all motives of the ad. Our analysis can therefore reveal which motives of the ad bring the most customers.

\begin{table}
\begin{center}
\caption{The mean absolute error (MAE), the root mean square error (RMSE), and the coefficient of determination (R$^2$) of the random forest on the training, validation, and testing samples for the cross-validation and the full and out-of-bag samples for the final model.}
\label{tab:treeError}
\begin{tabular}{llccc}
\toprule
Model & Sample & MAE & RMSE & R$^2$ \\ 
\midrule
Cross-Validation & Training   & 41.09 & 63.73 & 0.4693 \\ 
Cross-Validation & Validation & 51.34 & 78.82 & 0.1840 \\ 
Cross-Validation & Test       & 51.34 & 78.92 & 0.1839 \\ 
Final Model      & Full       & 40.57 & 62.71 & 0.4878 \\ 
Final Model      & Out-of-Bag & 50.42 & 77.74 & 0.2130 \\ 
\bottomrule
\end{tabular}
\end{center}
\end{table}

\section{Comparison to State-of-the-Art}

The recent study by \cite{Du2019} represents a pivotal reference in the field of immediate reaction estimation to television advertisements. Our methodology is primarily compared to this work. The model of \cite{Du2019} incorporates several key elements: the weekly hourly slope, the annual hourly base level, the search volume for a control keyword, and the audience size exposed to the advertisement, inclusive of both the focal company and its competitors. Notably, they account for a delay in advertising effects of up to nine minutes for the company's own advertisements. However, the interaction between control keyword searches and the TV advertising effect is still a subject of debate. A significant limitation in the study of \cite{Du2019} is the exclusion of periods with highly influential TV advertisements, such as during the Super Bowl.

The model of \cite{Du2019} features the slope for each hour of the week (resulting in $24 \times 7$ variables), the base level for each hour of the year ($365 \times 24$ variables per year), and the audience size exposed to the advertisement, considering delays up to nine minutes. The model also utilizes data on competitors’ advertisements and control keyword searches. Since we lack access to such data, we apply a simplified version of their model to our dataset. The approach of \cite{Du2019} introduces computational complexities, particularly in handling matrices with nearly 9,000 input variables.

\begin{figure}
\begin{center}
\includegraphics[width=0.9\textwidth]{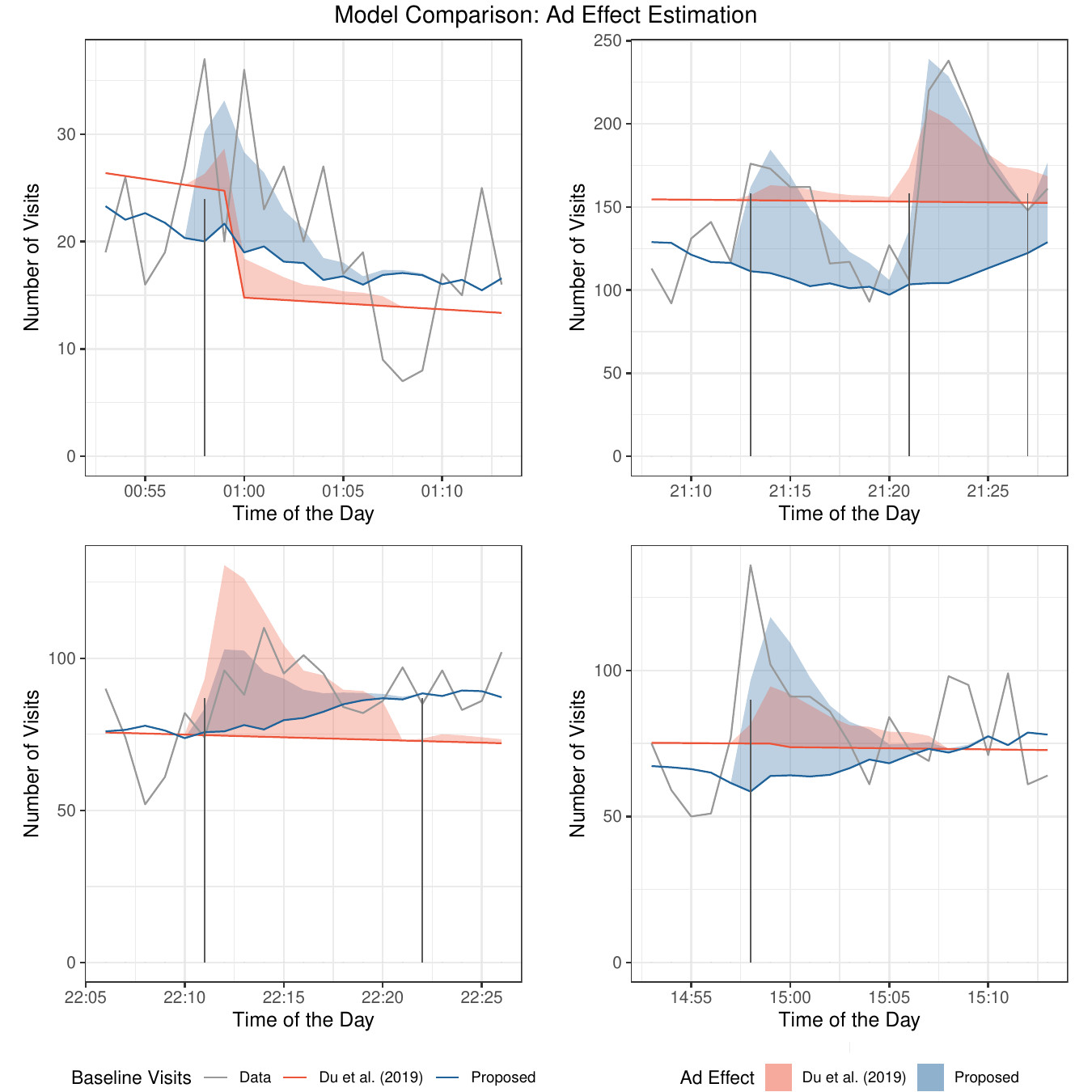}
\caption{Comparison of the estimated ad effects in specifically challenging situations. The grey bars are the location of the ads.}
\label{fig:compare}
\end{center}
\end{figure}

Another significant drawback of this method is its inability to accurately reflect rapid temporal fluctuations, such as the variation in visits within a single hour, particularly evident during evening times. This limitation is further compounded by the model's failure to ensure continuous diurnality estimation, potentially leading to significantly biased effect estimations, as illustrated in the top left section of Figure \ref{fig:compare}.

While \cite{Du2019} presume a linear correlation between the advertising effect and audience size, our study diverges from this assumption. We recognize that a linear relationship between audience size and effect is an oversimplification, contrary to established marketing theories and the concept of diminishing returns (see \citealp{Hruschka1993}). Instead, our model focuses on the advertisement's position and seeks to estimate visit increases before establishing causality. The dependent estimates may result in an inaccurate reflection of true changes in visitation patterns. This inconsistency is evident in Figure \ref{fig:compare}, where the effects are sometimes overestimated (right side) and at other times underestimated (left side).

\section{Conclusion}
\label{sec:con}

We measure the immediate impact of TV ads on website traffic. The results show that most people need about one minute to search for additional information after an ad is aired. However, approximately a third of ads have no impact at all. The most important factors driving the success of an ad are the time of the day, the TV channel, and the advertising motive of the ad.
 
Using predictions of the impact of an ad, a company can evaluate the additional value of the extra-paid qualities of the ad. For example, once the ad's date, time, and motive are known in the TV planning, the extra value of a premium position can be calculated. This approach can be used by marketers who want to optimize the TV ads buying process. To decide whether it is worth the investment, we would further need to know the average conversion rate on the website as well as the average value of a conversion. Moreover, it is possible in our framework to evaluate the most attractive advertising motive that generates the highest interest.

One direction for future research is to extend the specification of the point process modeling the website visits. The negative binomial distribution could account for overdispersion and an autocorrelation structure could be considered, similarly to e.g.\ \cite{Tomanova2021}. Another challenging direction is to study the short-term and long-term effects of TV advertising, as was done e.g.\ by \cite{Guitart2017}. Finally, a comprehensive comparison of our approach to \cite{Kitts2014}, \cite{Liaukonyte2015}, \cite{Fossen2017a}, \cite{Du2019}, and similar studies would bring a useful insight.

\section*{Funding}
\label{sec:fund}

This research was supported by the Internal Grant Agency of the Prague University of Economics and Business under grant F4/27/2020. Computational resources were supplied by the project ``e-Infrastruktura CZ'' (e-INFRA LM2018140) provided within the program Projects of Large Research, Development and Innovations Infrastructures.


\begin{thebibliography}{22}
\newcommand{\enquote}[1]{``#1''}
\providecommand{\natexlab}[1]{#1}
\providecommand{\url}[1]{\texttt{#1}}
\providecommand{\urlprefix}{}
\expandafter\ifx\csname urlstyle\endcsname\relax
  \providecommand{\doi}[1]{doi:\discretionary{}{}{}#1}\else
  \providecommand{\doi}{doi:\discretionary{}{}{}\begingroup
  \urlstyle{rm}\Url}\fi
\providecommand{\eprint}[2][]{\url{#2}}

\bibitem[{Behnamian \emph{et~al.}(2017)Behnamian, Millard, Banks, White,
  Richardson, and Pasher}]{Behnamian2017}
Behnamian A, Millard K, Banks SN, White L, Richardson M, Pasher J (2017).
\newblock \enquote{{A Systematic Approach for Variable Selection with Random
  Forests: Achieving Stable Variable Importance Values}.}
\newblock \emph{IEEE Geoscience and Remote Sensing Letters}, \textbf{14}(11),
  1988--1992.
\newblock ISSN 1545-598X.
\newblock \url{https://doi.org/10.1109/lgrs.2017.2745049}.

\bibitem[{Chandrasekaran \emph{et~al.}(2017)Chandrasekaran, Srinivasan, and
  Sihi}]{Chandrasekaran2017a}
Chandrasekaran D, Srinivasan R, Sihi D (2017).
\newblock \enquote{{Effects of Offline Ad Content on Online Brand Search:
  Insights from Super Bowl Advertising}.}
\newblock \emph{Journal of the Academy of Marketing Science}, \textbf{46}(3),
  403--430.
\newblock ISSN 0092-0703.
\newblock \url{https://doi.org/10.1007/s11747-017-0551-8}.

\bibitem[{Du \emph{et~al.}(2019)Du, Xu, and Wilbur}]{Du2019}
Du RY, Xu L, Wilbur KC (2019).
\newblock \enquote{{Immediate Responses of Online Brand Search and Price Search
  to TV Ads}.}
\newblock \emph{Journal of Marketing}, \textbf{83}(4), 81--100.
\newblock ISSN 0022-2429.
\newblock \url{https://doi.org/10.1177/0022242919847192}.

\bibitem[{Duff and Segijn(2019)}]{Duff2019}
Duff BRL, Segijn CM (2019).
\newblock \enquote{{Advertising in a Media Multitasking Era: Considerations and
  Future Directions}.}
\newblock \emph{Journal of Advertising}, \textbf{48}(1), 27--37.
\newblock ISSN 0091-3367.
\newblock \url{https://doi.org/10.1080/00913367.2019.1585306}.

\bibitem[{Fossen and Schweidel(2017)}]{Fossen2017a}
Fossen BL, Schweidel DA (2017).
\newblock \enquote{{Television Advertising and Online Word-of-Mouth: An
  Empirical Investigation of Social TV Activity}.}
\newblock \emph{Marketing Science}, \textbf{36}(1), 105--123.
\newblock ISSN 0732-2399.
\newblock \url{https://doi.org/10.1287/mksc.2016.1002}.

\bibitem[{Fossen and Schweidel(2019)}]{Fossen2019}
Fossen BL, Schweidel DA (2019).
\newblock \enquote{{Social TV, Advertising, and Sales: Are Social Shows Good
  for Advertisers?}}
\newblock \emph{Marketing Science}, \textbf{38}(2), 274--295.
\newblock ISSN 0732-2399.
\newblock \url{https://doi.org/10.1287/mksc.2018.1139}.

\bibitem[{Guitart and Hervet(2017)}]{Guitart2017}
Guitart IA, Hervet G (2017).
\newblock \enquote{{The Impact of Contextual Television Ads on Online
  Conversions: An Application in the Insurance Industry}.}
\newblock \emph{International Journal of Research in Marketing},
  \textbf{34}(2), 480--498.
\newblock ISSN 0167-8116.
\newblock \url{https://doi.org/10.1016/j.ijresmar.2016.10.002}.

\bibitem[{Guitart \emph{et~al.}(2020)Guitart, Hervet, and Gelper}]{Guitart2020}
Guitart IA, Hervet G, Gelper S (2020).
\newblock \enquote{{Competitive Advertising Strategies for Programmatic
  Television}.}
\newblock \emph{Journal of the Academy of Marketing Science}, \textbf{48}(4),
  753--775.
\newblock ISSN 0092-0703.
\newblock \url{https://doi.org/10.1007/s11747-019-00691-5}.

\bibitem[{Hastie \emph{et~al.}(2008)Hastie, Tibshirani, and
  Friedman}]{Hastie2008}
Hastie T, Tibshirani R, Friedman J (2008).
\newblock \emph{{The Elements of Statistical Learning}}.
\newblock Second Edition. Springer, New York.
\newblock ISBN 978-0-387-84857-0.
\newblock \url{https://doi.org/10.1007/978-0-387-84858-7}.

\bibitem[{Hruschka(1993)}]{Hruschka1993}
Hruschka H (1993).
\newblock \enquote{{Determining Market Response Functions by Neural Network
  Modeling: A Comparison to Econometric Techniques}.}
\newblock \emph{European Journal of Operational Research}, \textbf{66}(1),
  27--35.
\newblock ISSN 0377-2217.
\newblock \url{https://doi.org/10.1016/0377-2217(93)90203-y}.

\bibitem[{Joo \emph{et~al.}(2014)Joo, Wilbur, Cowgill, and Zhu}]{Joo2014}
Joo M, Wilbur KC, Cowgill B, Zhu Y (2014).
\newblock \enquote{{Television Advertising and Online Search}.}
\newblock \emph{Marketing Science}, \textbf{60}(1), 56--73.
\newblock ISSN 0025-1909.
\newblock \url{https://doi.org/10.1287/mnsc.2013.1741}.

\bibitem[{Kitts \emph{et~al.}(2014)Kitts, Bardaro, Au, Lee, Lee, Borchardt,
  Schwartz, Sobieski, and Wadsworth-Drake}]{Kitts2014}
Kitts B, Bardaro M, Au D, Lee A, Lee S, Borchardt J, Schwartz C, Sobieski J,
  Wadsworth-Drake J (2014).
\newblock \enquote{{Can Television Advertising Impact Be Measured on the Web?
  Web Spike Response as a Possible Conversion Tracking System for Television}.}
\newblock In \emph{Proceedings of the 8th International Workshop on Data Mining
  for Online Advertising},  1--9. Association for Computing Machinery.
\newblock ISBN 978-1-4503-2999-6.
\newblock \url{https://doi.org/10.1145/2648584.2648591}.

\bibitem[{Liaukonyte \emph{et~al.}(2015)Liaukonyte, Teixeira, and
  Wilbur}]{Liaukonyte2015}
Liaukonyte J, Teixeira T, Wilbur KC (2015).
\newblock \enquote{{Television Advertising and Online Shopping}.}
\newblock \emph{Marketing Science}, \textbf{34}(3), 311--330.
\newblock ISSN 0732-2399.
\newblock \url{https://doi.org/10.1287/mksc.2014.0899}.

\bibitem[{Molnar(2019)}]{Molnar2019}
Molnar C (2019).
\newblock \emph{{Interpretable Machine Learning: A Guide for Making Black Box
  Models Explainable}}.
\newblock lulu.com.
\newblock ISBN 978-0-244-76852-2.
\newblock \url{https://doi.org/10.1201/9780367816377-16}.
\newblock \urlprefix\url{https://christophm.github.io/interpretable-ml-book/}.

\bibitem[{Nelder and Mead(1965)}]{Nelder1965}
Nelder JA, Mead R (1965).
\newblock \enquote{{A Simplex Method for Function Minimization}.}
\newblock \emph{The Computer Journal}, \textbf{7}(4), 308--313.
\newblock ISSN 0010-4620.
\newblock \url{https://doi.org/10.1093/comjnl/7.4.308}.

\bibitem[{{Randall A. Lewis} and {David H. Reiley}(2013)}]{Lewis2013}
{Randall A Lewis}, {David H Reiley} (2013).
\newblock \enquote{{Down-to-the-Minute Effects of Super Bowl Advertising on
  Online Search Behavior}.}
\newblock In {Association for Computing Machinery} (Ed.), \emph{Proceedings of
  the 14th ACM conference on Electronic commerce},  639--656.
\newblock ISBN 978-1-4503-1962-1.
\newblock \url{https://doi.org/10.1145/2482540.2482600}.

\bibitem[{Stacy(1962)}]{Stacy1962}
Stacy EW (1962).
\newblock \enquote{{A Generalization of the Gamma Distribution}.}
\newblock \emph{The Annals of Mathematical Statistics}, \textbf{33}(3),
  1187--1192.
\newblock ISSN 0003-4851.
\newblock \url{https://doi.org/10.2307/2237889}.

\bibitem[{Tirunillai and Tellis(2017)}]{Tirunillai2017}
Tirunillai S, Tellis GJ (2017).
\newblock \enquote{{Does Offline TV Advertising Affect Online Chatter?
  Quasi-Experimental Analysis Using Synthetic Control}.}
\newblock \emph{Marketing Science}, \textbf{36}(6), 862--878.
\newblock ISSN 0732-2399.
\newblock \url{https://doi.org/10.1287/mksc.2017.1040}.

\bibitem[{Tomanov{\'{a}} and Hol{\'{y}}(2021)}]{Tomanova2021}
Tomanov{\'{a}} P, Hol{\'{y}} V (2021).
\newblock \enquote{{Clustering of Arrivals in Queueing Systems: Autoregressive
  Conditional Duration Approach}.}
\newblock \emph{Central European Journal of Operations Research},
  \textbf{29}(3), 859--874.
\newblock ISSN 1435-246X.
\newblock \url{https://doi.org/10.1007/s10100-021-00744-7}.

\bibitem[{Wilbur(2016)}]{Wilbur2016}
Wilbur KC (2016).
\newblock \enquote{{Advertising Content and Television Advertising Avoidance}.}
\newblock \emph{Journal of Media Economics}, \textbf{29}(2), 51--72.
\newblock ISSN 0899-7764.
\newblock \url{https://doi.org/10.1080/08997764.2016.1170022}.

\bibitem[{Wright(2015)}]{Wright2015}
Wright SJ (2015).
\newblock \enquote{{Coordinate Descent Algorithms}.}
\newblock \emph{Mathematical Programming}, \textbf{151}(1), 3--34.
\newblock ISSN 1436-4646.
\newblock \url{https://doi.org/10.1007/s10107-015-0892-3}.

\bibitem[{Zhang \emph{et~al.}(2017)Zhang, Trusov, Stephen, and
  Jamal}]{Zhang2017}
Zhang Y, Trusov M, Stephen AT, Jamal Z (2017).
\newblock \enquote{{Online Shopping and Social Media: Friends or Foes?}}
\newblock \emph{Journal of Marketing}, \textbf{81}(6), 24--41.
\newblock ISSN 0022-2429.
\newblock \url{https://doi.org/10.1509/jm.14.0344}.

\end{thebibliography}

\end{document}